\documentclass[a4paper,12pt]{article}

\title{Gravity-induced four-fermion contact interaction implies gravitational intermediate W and Z type gauge bosons}

\author{Jens Boos\,\footnote{~Corresponding author, E-mail:
    \href{mailto:jboos@perimeterinstitute.ca}{jboos@perimeterinstitute.ca}}
  ${}^{,1,2}$ and Friedrich W.\ Hehl\,\footnote{~E-mail:
    \href{mailto:hehl@thp.uni-koeln.de}{hehl@thp.uni-koeln.de}}${}^{~,3,4}$ \\
    {\small ${}^1$ Perimeter Inst.\ Theor.\ Physics, Waterloo, ON N2L 2Y5, Canada} \\[-5pt]
	{\small ${}^2$ Dept.\ Physics \& Astron., Univ.\ of Waterloo, Waterloo, ON N2L 3G1, Canada}\\[-5pt]
    {\small ${}^3$ Inst.\ Theor.\ Physics, Univ.\ of Cologne, 50923 K\"oln, Germany} \\[-5pt]
    {\small ${}^4$ Dept.\ Physics \& Astron., Univ.\ of Missouri, Columbia, MO 65211, USA} }

\date{15 September 2016}

\usepackage[english]{babel}
\usepackage{amsmath}
\usepackage{amsbsy}
\usepackage{amstext}
\usepackage{amscd}
\usepackage{amsxtra}
\usepackage{amsopn}
\usepackage[onehalfspacing]{setspace}
\usepackage{tensor}
\usepackage{mathrsfs}
\usepackage{relsize}
\usepackage{amsfonts}
\usepackage{MnSymbol}
\usepackage[makeroom]{cancel}
\usepackage{selinput}
\usepackage{bbm}

\usepackage{youngtab}

\newcommand{\dd}{\mbox{d}}

\usepackage{geometry}
\usepackage{parskip}

\usepackage[colorlinks=true,citecolor=green,linkcolor=red,filecolor=cyan,urlcolor=magenta]{hyperref}

\usepackage{color}
\definecolor{mygray}{rgb}{0.5,0.5,0.5}

\usepackage{listings} \definecolor{fadedred}{rgb}{0.6,0.0,0.0}
\definecolor{fadedblue}{rgb}{0.0,0.0,0.6}

\begin{document}

\maketitle

\begin{abstract}
  Coupling fermions to gravity necessarily leads to a
  non-renormalizable, gravitational four-fermion contact
  interaction. In this essay, we argue that augmenting the
  Einstein--Cartan Lagrangian with suitable kinetic terms quadratic in
  the gravitational gauge field strengths (torsion and
    curvature) gives rise to new, massive propagating gravitational
  degrees of freedom.  This is to be seen in close analogy to Fermi's
  effective four-fermion interaction and its emergent W and Z bosons.

\hfill {\small \textit{file:
      EssayIJTP2016\_5.tex, 15 Sep 2016, jb, fwh} }
\end{abstract}

\vspace{20em}

\pagebreak

{\it The idea that spin gives rise to torsion should not be regarded
  as an ad hoc modification of General Relativity. On the contrary, it
  has a deep group theoretical and geometric basis. If history had
  been reversed and the spin of the electron discovered before 1915, I
  have little doubt that Einstein would have wanted to include torsion
  in his original formulation of General Relativity. On the other
  hand, the numerical differences which arise are normally very small,
  so that the advantages of including torsion are entirely
  theoretical.}
\begin{flushright}
--- Dennis W.\ Sciama, priv.\ comm.\ (1979)
\end{flushright}

\section{Introduction}

Physics thrives, whenever concepts of different disciplines are
combined to find something new. One may think of electromagnetism, the
archetype of classical field theory, which in many ways served as a
precursor for the more elaborate theory of General Relativity. Up to
today, physicists use and employ similarities between these two
theories to learn something new about the other, and in the light (or
sound) of the gravitational wave detection GW150914 this becomes even
more apparent. Other examples are renormalization and massive gauge
theories, enriched by physical understanding via the condensed matter
phenomena of block spin coarse-graining and spontaneous symmetry
breaking, respectively.

Undoubtedly, the quantum nature of gravity has been a very puzzling
question of both the twentieth and twenty-first century. While we do
not attempt to settle this issue here, we would like to employ an
analogy from particle theory, that could perhaps open a fruitful and
interesting direction for future research.

In 1933, Fermi proposed a phenomenological model of a four-fermion
interaction that could potentially describe the $\beta$ decay. Due to
the negative mass dimension coupling constant, $G_{\text{F}} \simeq
1.17 \times 10^{-5} \, \text{GeV}^{-2} \, (\hbar c)^3$, which is
equivalent to a typical interaction range of $l_{\text{Fermi}}\approx
10^{-20} \, \text{m}$, this theory is non-renormalizable and should
be viewed as an effective field theory instead, as it turned out
later. However, a massive gauge boson propagator can be approximated
as
\begin{align}
  \frac{-i}{p^2 - M^2 + i\epsilon}\left( g{}_{ij} - \frac{p{}_i
      p{}_j}{M^2} \right) \quad \approx \quad \frac{i g{}_{ij}}{M^2}
  \quad \propto \quad i \, G{}_\text{F} \, g{}_{ij} \, ,
\end{align}
provided that $p \ll M$. Here p is the momentum of the particle, M its
mass, $g_{ij}$ the metric of spacetime (here $i,j,...$ are coordinate
indices running from $0$ to $3$). Hence, an effective field theory,
even though non-renormalizable, can give important hints to the more
accurate physics of nature (in this case, the underlying massive SU(2)
gauge structure of the weak interaction).

Let us now turn back to gravity---at this point, we would like
to quote John L.\ Synge, who once said ``Newton successfully wrote
apple = moon, but you cannot write apple = neutron.'' This is true due
to the intrinsically fermionic character of matter. As a historical
remark, at times of the discovery of the field equation of General
Relativity, spin was still not known. It is hence not surprising that
General Relativity fails taking intrinsic angular momentum (the
classical analogue of spin) into account: it only couples to the
symmetric Hilbert energy-momentum tensor, which is blind to the spin current.

Hence Riemannian geometry, the
geometrical arena of General Relativity, is not large enough to
accommodate spin. However, it has long been shown that the
Einstein--Cartan theory, naturally endowed with a Riemann--Cartan
geometry allowing for non-vanishing torsion $T_{\alpha\beta}{}^i$, is
a consistent candidate theory for coupling spin to gravity; here
$\alpha,\beta,...$ are frame indices taking the values $\hat 0,\hat
1,\hat 2,\hat 3$. Therein, energy-momentum sources curvature, and spin
sources torsion. As it turns out, the tetrad 1-form
$e{}_i{}^\alpha \dd x{}^i$ (coframe) and the connection 1-form
$\Gamma{}_i{}^{\alpha\beta} \dd x^i$ are now independent translational
and rotational gauge fields, respectively. Together they
incorporate local Poincar\'e invariance \cite{Blagojevic:2013}. Their
field equations read
\begin{align}\label{fieldeqs}
  \text{Ric}{}_\alpha{}^i - \frac 12 R e{}^i{}_\alpha + \Lambda
  e{}^i{}_\alpha = \kappa\, \mathfrak{T}{}_\alpha{}^i, \qquad
  T{}_{\alpha\beta}{}^i + \delta{}^i_{\!\alpha} T{}_{\beta\gamma}{}^\gamma
  - \delta{}^i_{\!\beta} T{}_{\alpha\gamma}{}^\gamma = \kappa
  \,\mathfrak{S}{}_{\alpha\beta}{}^i ,
\end{align}
where $e \, \mathfrak{T}{}_\alpha{}^i := \delta \mathcal{L} / \delta
e{}_i{}^\alpha$ is the canonical energy-momentum of matter, $e \,
\mathfrak{S}{}_{\alpha\beta}{}^i := \delta \mathcal{L} / \delta
\Gamma{}_i{}^{\alpha\beta}$ is the canonical spin current of matter,
and $e := \text{det}\left( e{}_i{}^\alpha \right)$. On the other hand,
the symmetric Hilbert energy-momentum tensor of General
Relativity is given by $\sqrt{-g} \, t{}^{ij} := 2 \delta \mathcal{L}
/ \delta g{}_{ij}$, with $g := \text{det}\left( g{}_{ij} \right)$. In
order to establish the difference of Einstein--Cartan theory as
compared to General Relativity, one can write (see
\cite{Blagojevic:2013} and also Pop\l{}awski \cite{Poplawski:2011jz},
Magueijo \textit{et al}.\ \cite{Magueijo:2012ug}, and Khriplovich and
Rudenko \cite{Khriplovich:2013tqa})
\begin{align} {}^{\text{EC}}t{}^{ij} := t{}^{ij} + \kappa \left[ -4
    \mathfrak{S}^{ik}{}_{[l} \mathfrak{S}{}^{jl}{}_{k]} - 2
    \mathfrak{S}{}^{ikl} \mathfrak{S}{}^{j}{}_{kl} +
    \mathfrak{S}{}^{kli} \mathfrak{S}{}_{kl}{}^j + \frac 12 g{}^{ij}
    \left( 4 \mathfrak{S}{}_m{}^k{}_{[l} \mathfrak{S}{}^{ml}{}_{k]} +
      \mathfrak{S}{}^{mkl} \mathfrak{S}{}_{mkl} \right) \right]
  , \label{eq:spinsquared}
\end{align}
where $\kappa = 8\pi G/c^4$ is Einstein's gravitational constant. From
the perspective of General Relativity, we integrated out torsion and
thereby created an additional spin-spin contact interaction, or a
\emph{gravitational four fermion interaction} beyond General
Relativity. The ``coupling constant'' of this four-fermion interaction
is indeed of mass dimension $-2$, that is, $\kappa = \hbar / (c^3
M^2)$, where $M = \sqrt{\hbar c/(8\pi G)} = 2.4 \times 10^{18} \,
\text{GeV}$ is the reduced Planck mass. Moreover, this four-fermion
interaction in general contains parity-even
$\overline{\Psi}\gamma{}^\alpha\Psi$ and parity-odd
$\overline{\Psi}\gamma{}^5\gamma{}^\alpha\Psi$ pieces (where $\Psi$ is
the spinor field, $\overline{\Psi}$ its Dirac adjoint, and
$\gamma{}^\alpha$ are the Dirac matrices), in close analogy to the
electroweak scenario where the analogy began in the first place.

Of the two field equations in \eqref{fieldeqs}, the first one
corresponds to translational invariance (generator $P{}_\alpha$) and
energy-momentum, the second one to Lorentz invariance (generator
$J{}_{\alpha\beta}$) and spin angular momentum. These facts are also
manifest in the study of the {\it gravitational phase shift}
integrated along an infinitesimal closed loop $\gamma$ bordering the
area element $\text{d}\sigma{}^{ij}$. This yields, as shown by Anandan
\cite{Anandan:1994tx},
\begin{equation}
  \Phi_\gamma=1-\frac i2 \left(T{}_{ij}{}^\alpha P{}_\alpha 
    + R{}_{ij}{}^{\alpha\beta}J{}_{\alpha\beta}\right)\text{d}\sigma{}^{ij}\,.
  \label{eq:pgt-holonomy}
\end{equation}
Torsion is related to translation and curvature to Lorentz
transformations, as a geometrical interpretation of the
Riemann--Cartan geometry would suggest, and hence the holonomy
of the closed loop $\gamma$ is a Poincar\'e transformation.

\section{Gravitational four-fermion interaction}

Let us now estimate at which energy scales this interaction, should it
be realized in nature, becomes relevant. Since it is related to
particles that carry spin, we demand that the quantities on the
right-hand side of Eq.\ \eqref{eq:spinsquared} are of the same order
of magnitude, that is,
\begin{align}
  t{}^{ij} \approx \kappa {\left( \mathfrak{S}^2 \right)}^{ij}
  , \label{eq:relevance}
\end{align}
see the phenomenological oriented discussion of Ni
\cite{Ni:2009}. Given a certain number density $n$ of fermions with
mass $m$, the mass density is $\rho = m \, n$. Next, we need to
estimate the spin density $s$. In standard equilibrium configurations
and in the absence of external magnetic fields, spins tend to average
out, and hence $s \in \mathcal{O}\left(\hbar\right)$. In order for the
gravitational four-fermion interaction to become relevant, we either
need to assume large magnetic fields or special phases of matter, like
a ferromagnetic phase, for instance. Both criteria can be met in
extreme situations: neutron stars are known to have extremely strong
magnetic fields of the order of $10^4 \dots 10^{11}\,\text{T}$, and
anomalous phases like the low temperature A phase of ${}^3$He indeed
have a macroscopic net spin, see the books by Vollhardt and W\"olfle
\cite{Vollhardt:2013} and Volovik \cite{Volovik:2003}. It is therefore
conceivable that there are, indeed, both astrophysical and
cosmological scenarios where the spin density can be approximated by
$s \approx \hbar \, n$.

Substitution into Eq.\ \eqref{eq:relevance} yields the critical or
\emph{Einstein--Cartan number density} $n_\text{EC} \approx
m/(\kappa\hbar)^2$. Defining the reduced Compton wavelength of the
fermion under consideration, $\lambda{}_\text{Compton} := \hbar/(mc)$,
one finally has for the critical Einstein--Cartan density
\begin{align}
  \rho{}_\text{EC} = m\, n{}_\text{EC} =
  \frac{m}{\lambda_\text{Compton}\,\ell{}_\text{Planck}^2} .
\end{align}
For a typical nucleon, $m \approx 1\,\text{GeV}/(c^2)$,
 and hence $\rho_\text{EC} \approx 10^{59} \,
  \text{kg}/\text{m}^3$, which is much smaller than the reduced
Planck density of $\rho{}_\text{Planck} = 10^{96} \,
  \text{kg}/\text{m}^3$ at the big bang. For comparison, a typical
nuclear density is $\rho_\text{nucl} = 10^{18} \,
\text{kg}/\text{m}^3$. Analogously, the \textit{Einstein--Cartan
  length} scale $\ell{}_\text{EC} = \left(\lambda{}_\text{Compton} \,
  \ell{}_\text{Planck}^2\right)^{1/3} \approx 10^{-29} \, \text{m}$
is seven orders of magnitude larger than the reduced Planck scale
$\ell{}_\text{Planck} \approx 10^{-36}\,\text{m}$.

PLANCK data indicate \cite{Mukhanov:2013} that General Relativity can
be verified to scales of $\approx 10^{-28}\,\text{m}$. Hence,
noticeable effects due to Einstein--Cartan corrections can be expected
to emerge soon, if present.

\section{Liberating the gravitational W and Z bosons}

Similar as in Fermi's original contact interaction, this gravitational
four-fermion contact interaction is probably not the end of the
story. Contact interactions are unphysical, since they are mediated by
some Heaviside potential, differentiations of which produce infinite,
delta function-like forces.

Like in the electroweak case, we may now introduce
additional, massive short-range degrees of freedom by modifying the
Lagrangian of the Einstein--Cartan theory. However, there are
two key ingredients that allow us to make an almost unique choice: (i)
the coupling constant is an inverse mass squared, hence we know that
the additional degree of freedom has to correspond to massive
bosons. (ii) The Einstein--Cartan theory results from a
gauge approach to gravity, hence there is a natural way to
include kinetic terms of the gauge potentials $e{}_i{}^\alpha$
and $\Gamma{}_i{}^{\alpha\beta}$: the squares of their respective
curvatures, that is, torsion $T{}_{\alpha\beta}{}^i$ (translational
curvature) and curvature $R{}_{\alpha\beta}{}^{ij}$ (rotational
curvature), correspond to the only gauge-invariant kinetic terms
allowed.

Demanding that the field equations for coframe and connection be
linear in second derivatives, one can construct the following type of
extended Einstein--Cartan Lagrangian:
\begin{align}
  \mathcal{L}{}^\text{ext.}_\text{EC} \quad \sim \quad e \, \left[
    \frac{1}{(\mu\ell{}_\text{Planck})^4} + \frac{1}{L^2} \, \left(
      T{}_{\alpha\beta}{}^i\, T{}^{\alpha\beta}{}_i + \frac{1}{\chi}
      R{}_{\alpha\beta}{}^{ij}\, e{}_i{}^\alpha e{}_j{}^\beta \right)
    + \frac{1}{k} R{}_{\alpha\beta}{}^{ij} R{}^{\alpha\beta}{}_{ij}
  \right] \label{eq:lagrangian}
\end{align}
Here, $\mu$, $\chi$, and $k$ are new, dimensionless parameters, and
$L$ is a length scale such that $\chi L^2 =
\ell{}_\text{Planck}^2$. For an explicit form of the Lagrangian
\eqref{eq:lagrangian}---including parity-even and parity-odd
terms---see Eq.\ (5.13) in Blagojevi\'c and Hehl
\cite{Blagojevic:2013}. Lagrangians like the above have been studied,
see Sezgin and van Nieuwenhuizen \cite{Sezgin:1979zf} or Kuhfuss and
Nitsch \cite{Kuhfuss:1986rb}, and found to be ghost-free under certain
conditions; see also the recent work of Shie \textit{et al}.\
\cite{Shie:2008ms} or Bjorken \cite{Bjorken:2010qx} in the context of
cosmology. As a finger exercise, we recently studied quadratic
curvature terms in the context of exact solutions of Einstein's
equations \cite{Boos:2014hua}.

The coupling constant of the curvature-squared piece is dimensionless,
just like one would expect for Yang--Mills theory, and we call the
propagating degrees of freedom liberated by this kinetic term
``tordions.'' As it turns out, they are also massive, $m =
\sqrt{k}/\ell{}_\text{Planck}$, and for the interaction range to be of
order $\ell{}_\text{EC}$, as argued above, one has $k \approx
(\ell{}_\text{Planck}/\ell{}_\text{EC}){}^2 \approx 10^{-14}$. Hence,
we arrived at a theory with a weak, Einstein gravity sector mediated
by $e{}_i{}^\alpha$ (gravitons), and a strong, massive Yang--Mills
sector mediated by $\Gamma{}_i{}^{\alpha\beta}$ (``gravitational
W and Z type bosons'').

\section{Conclusion}

In this essay, we have argued that the gravitational four-fermion
interaction can possibly be described as the effective field theory
limit of a curvature-squared Lagrangian in the framework of Poincar\'e
gauge theory of gravity. We leave a more detailed analysis for the future.

The new gravitational degrees of freedom are of a geometric origin rooted
in the Poincar\'e group.  Their similarities to the electroweak W and Z
bosons are purely based on effective field theory considerations, and we
cannot see any geometric relation between torsion and electromagnetism,
see, however, Pop\l{}awski \cite{Poplawski:2009wd}, e.g., and references
therein.

\textit{Note added in proof:} The recent paper of Donoghue
\cite{Donoghue:2016vck} seems to have a similar aim as ours. Donoghue
first reiterates the standard point of view of Poincar\'e gauge theory,
compare \cite{Blagojevic:2013}. What he calls `the constraint of
metricity for the vierbeins,' translates into our language as the
vanishing of Cartan's torsion tensor. The torsion itself is introduced
by Donoghue in his Eq.~(29). A link variable for torsion is suggested
by comparing Donoghue's Eq.~(32) with our Eq. \eqref{eq:pgt-holonomy},
namely as proportional to $P_a e_\mu{}^a$.

Donoghue has shown that the beta function of an SO(1,3) gauge theory
is negative. Interpreting this as the Lorentz sector of Poincar\'e
gauge theory, it implies that possibly new propagating degrees of
freedom (tordions, or ``gravitational W and Z bosons'') are confined,
similar as the fermionic quarks in quantum chromodynamics. Due to the
positivity of mass, however, this confinement might be of different
origin than the screening mechanism encountered in particle theory.
Nevertheless, this is a highly interesting observation and should be
compared with the approach advocated here, since both mechanisms
render the new degrees of freedom irrelevant for large scale physics.

\section*{Acknowledgments}
Research at Perimeter Institute is supported by the Government of
Canada through Industry Canada and by the Province of Ontario through
the Ministry of Research and Innovation.

\begin{singlespace}

\end{singlespace}

\end{document}